\documentclass[10pt,a4paper]{article}

\usepackage{times}

\usepackage{epsfig}
\usepackage{epstopdf} 

\usepackage{fancyhdr}

\usepackage[lmargin=2.5cm, rmargin=2.5cm,tmargin=3.50cm,bmargin=3.50cm]{geometry}
\usepackage{indentfirst}

\def\e{\begin{equation}}
\def\f{\end{equation}}
\def\_#1{{\bf #1}}

\def\.{\cdot}

\begin{document}

\title{\large \textbf{Waveguide Bandpass Filters made of Thick Complementary Small Resonators}}
%
\def\affil#1{\begin{itemize} \item[] #1 \end{itemize}}
\author{\normalsize \bfseries \underline{L. M. Pulido-Mancera} and J. D. Baena}
%
\date{}
\maketitle
\thispagestyle{fancy} 
\vspace{-6ex}
\affil{\begin{center} Universidad Nacional de Colombia, Physics Department, Bogot\'a, Colombia\\
lmpulidom@unal.edu.co\\
 \end{center}}

\begin{abstract}
We explore the potential use of complementary split ring resonators (csrr's) and complementary spirals of two and three turns (csr2 and csr3) to design very compact waveguide filters. We demonstrate its passband response in an experimental and computational form. Besides, it is shown that the design process can be highly simplified by means of a new waveguide model that reduce full wave numerical simulations. Thus, it is possible to design waveguide filters with lengths equal to the thickness of a metallic sheets, leading to an improvement in the design of miniaturized waveguide filters. 
\end{abstract}

\section{Introduction}
The split ring resonator (srr) was first introduced by Pendry et al. (1999) in order to produce artificial media with a strong magnetic response at microwaves and radio frequencies \cite{pendry}. Its complementary structure, the csrr recently reported by \cite{Babinet-Baena}, have been inspired on the Babinet's principle and, similar to the SRRs, it is an electrically small particle that exhibits a quasi-static resonance. A 2D array of csrrs behaves as a passband filter whose resonance frequency depends on its geometrical parameters only. Similarly, for waveguide applications it allows us to replace the well-known resonant cavities coupled by circular irises in waveguide filters, resulting in a compact design \cite{noelia}. 

The design process of infinitely thin csrrs and similar geometries like spirals, has been developed by means of equivalent LC-Circuit Models in order to predict the resonance frequency: The intrinsic circuit model for the csrr (dual of the SRR model) has a parallel combination of two inductances connecting the inner disk to the ground and the capacitance of a disk of radius $r$ surrounded by a ground plane at a distance of its edge. Likewise, equivalent circuit models for planar csr2 and csr3 can be found in \cite{LC-Circuit-Baena}.
However, when the thickness increases, duality given by Babinet's principle is no longer valid and different approaches to predict the resonance in terms of the thickness are needed \cite{thick-screens}.

\section{Waveguide model for thick csrr}

\begin{figure}
\centering \includegraphics[scale=0.8]{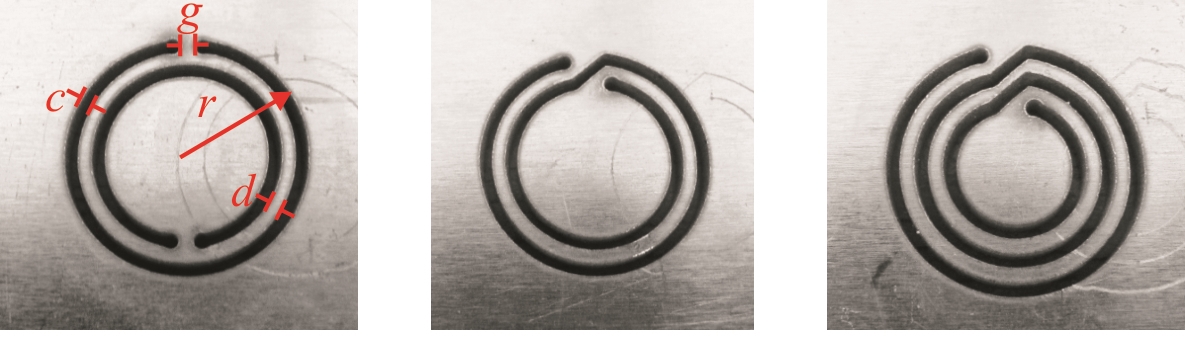} 
\caption{Resonators used for metasurfaces design. a) csrr, b)csr2, c) csr3. The geometric parameters for all the geometries are $r= 9.0 mm$, $c=d=1.0mm$, $g= 2.9mm$ and $h$ varies from 1.2 mm to 5.8 mm. The thinnest layer is $h=1.2$mm and the thicker layer is $h=2.6$mm} \label{fig1}
\end{figure}

Let us consider metallic plate with the csrr geometry etched on it. A frontal view of the resonator with its geometrical parameters are shown in Figure \ref{fig1}a). The same geometrical parameters were used for the csr2 (Fig. \ref{fig1}b) and the csr3 (Fig. \ref{fig1}c). The resonator is excited by a harmonic magnetic field towards the $x$ direction, corresponding to the $TE_{10}$ mode of the rectangular waveguide. In our model, when the surface is thick, each slit of the csrr (csr2 or csr3) can be seen as a rectangular waveguide bended with the shape of C or spiral respectively.  Hence, as a waveguide, the lowest frequency that allows maximum transmission is the cutoff frequency $f_{0}$ of the mode $TE_{10}$, when $\lambda_{0}=2L$ with $\lambda_{0}$ the resonance wavelength and $L$ the length of the slit. The second resonance $f_{1}$ (associated to $\lambda_{1}$) appears when $\lambda_{1}=L$ is satisfied. For each prototype, $L$, $f_{0}$, and $f_{1}$ predicted of our simplified model are shown in Table \ref{table1}.

\begin{table}
\centering
\caption{Resonances expected with the waveguide model for thick resonators.}
\begin{tabular}{|c|c|c|c|}
\hline
Properties & csrr & csr2 & csr3\\ \hline
Length /mm & 52.40  & 94.34 & 125.61 \\ \hline
$f_{0}$ / GHz & 2.86 & 1.58 & 1.19 \\ \hline
$f_{1}$ / GHz & 5.72 & 3.16 & 2.38 \\ \hline
\end{tabular}
\label{table1}
\end{table}

\section{Experimental Setup}
Experiments were carried out by placing the metallic resonator layers shown in Fig. \ref{fig1} between two waveguides connected through the flange. For each geometry, a set of metallic layers were placed one behind each other between two standard WR340 waveguide of 86 $\times$ 43 mm$^{2}$ cross sectional area and cutoff frequency at 1.736 GHz. This area is 16 times bigger than the resonator's area which allow us to find the resonance frequency of a single resonator rather than the resonance of its images produced by the electric walls of the WR340 waveguide. 

\begin{figure}[h!]
\centering \includegraphics[width=0.9\textwidth]{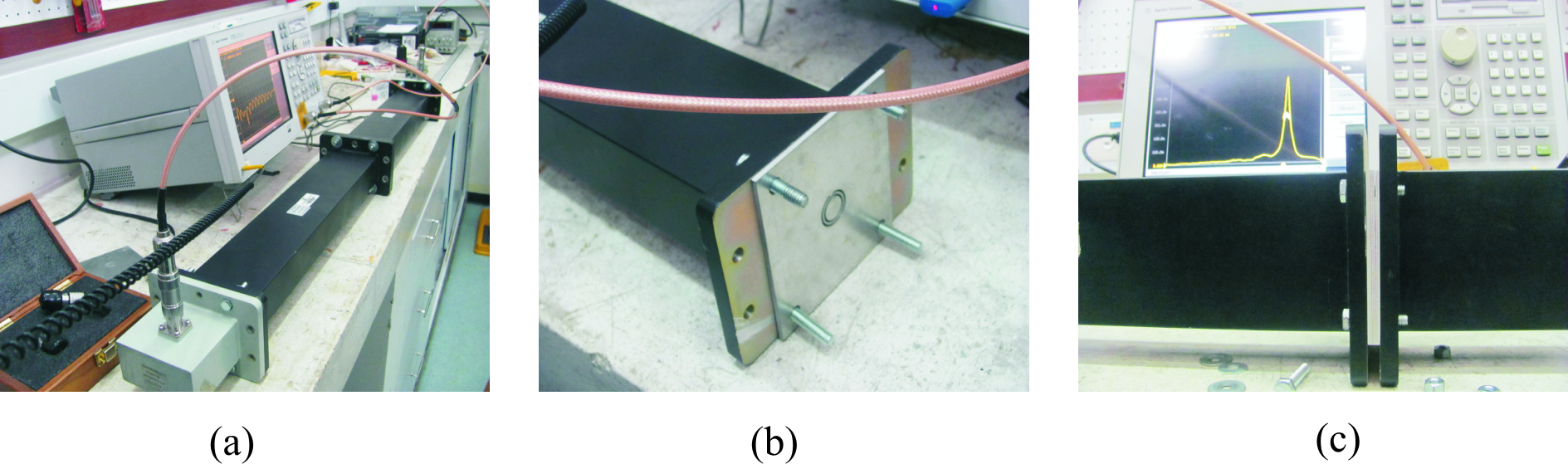}   
\caption{Experimental Setup. a)Setup to measure the transmission band when the resonator is placed between two waveguides WR340. b) Orientation of the re resonator in front of the aperture to get the proper excitaion mode. c) Equivalent thick resonator} \label{fig2}
\end{figure}

Full two port calibration up to the coaxial port was performed before the measurements. Each WR340 is connected at the and to coaxial adapters, which are used to excite and detect the signal as shown in Fig. \ref{fig2}a). The vector network analyzer Agilent E5062 is used to measure the $|S_{21}|$ and $|S_{11}|$ parameters every time that a new layer is inserted between the waveguides: Fig. \ref{fig2}b) shows how to create a thicker resonator, Fig. \ref{fig2}c). Considering that we are focused in the passband response of these structures, in this paper we will present the results for the $S_{21}$ parameter of the geometries whose resonance associated to $\lambda_{0}$ or $\lambda_{1}$ appears. 
 
\section{Results}
\begin{figure}[h!]
\centering \includegraphics[width=0.75\textwidth]{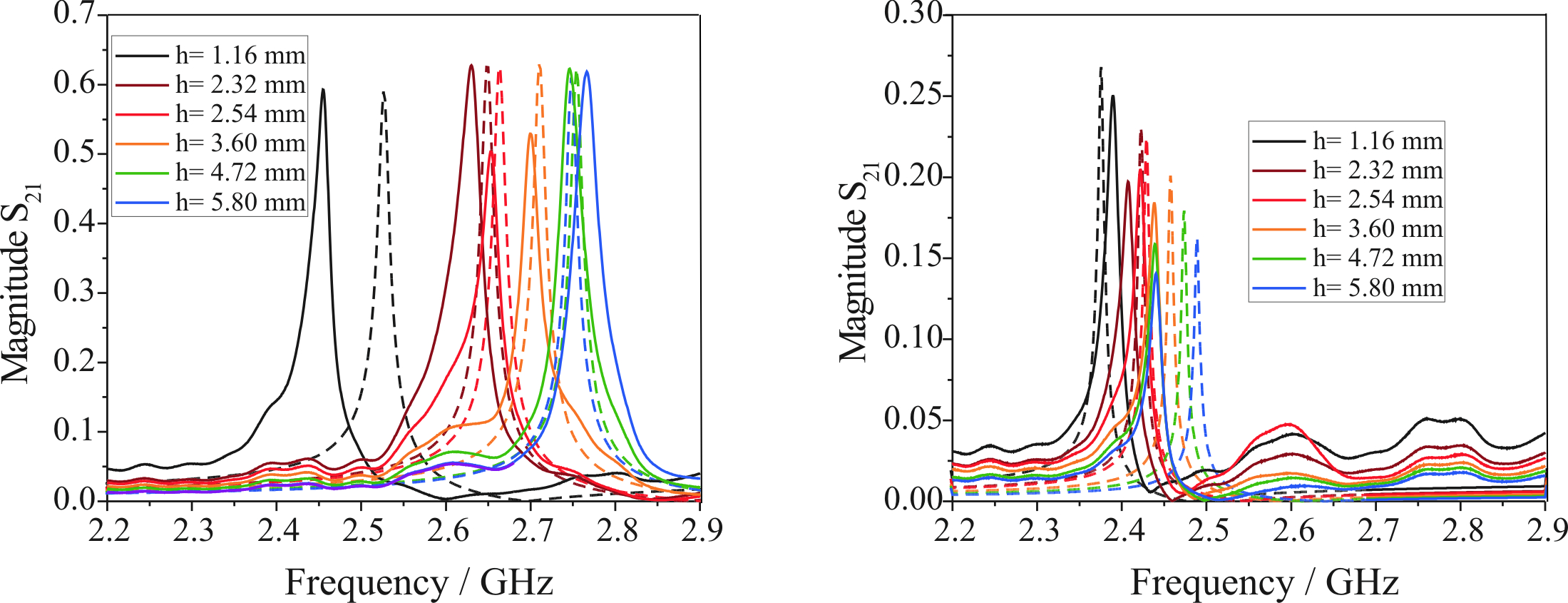}  
\caption{Results from experiments and simulations varying the thickness of the filter. a) $S_{21}$ for csrr (first resonance $f_{0}$), b) $S_{21}$ for csr2 (no resonance), C) $S_{21}$ for csr3 (second resonance $f_{1}$).}
\label{fig3}
\end{figure}

The results for the experimental setup of Fig \ref{fig2} relating to csrr, csr2, and csr3 are shown in Fig \ref{fig3}. In all cases, the $|S_{21}|$ and $|S_{11}|$ parameters were measured and compared with full wave numerical simulation using \textit{CST- Microwave Studio} software within the frequency range [2 - 3 GHz]. 
It may be observed that the amplitude of peaks found in the $S-$parameters is only slightly greater in simulations than in the experiment, and this amplitude remains approximately constant despite the resonator becomes thicker and higher losses are expected. The differences in resonance frequency $f_{0}$ are less than 10$\%$ for the thinnest layer ($h =$ 1.16) and less than $1\%$ when the surface is thick. See Fig \ref{fig3}.
On the other hand, it can be seen that when the thickness of the filter increases, $f_{0}$ varies only 350 MHz, approaching constant value of 2.82 GHz, which corresponds to the frequency of the fundamental mode shown in Table \ref{table1}. Thus, we have verified experimentally our previously waveguide model for thick csrrs based resonators. In the case of the spiral of two turns, csr2, it had been previously shown that the relation between the resonance of the csrr and csr2 of equal dimensions satisfies $f_{0}^{csrr} = 2f_{0}^{csr2}$ \cite{AP-S2013}. Thus, it is expected no resonance within the range of frequencies of the WR340. Similarly, for the spiral of three turns (csr3), $f_{0}^{csrr} = 2\sqrt{2}f_{0}^{csr3}$ is satisfied. Again, $f_{0}$ does not appear within the range of frequencies of the WR340. Instead, it can be seen (See Fig \ref{fig3}) the first high order resonance $f_{1}$ of Table \ref{table1}. Therefore, our model not only predicts the first resonance as the cutoff frequency of the equivalent waveguide, but also predicts higher-order resonances. However, in this last case the losses are greater, so that the filter csr3 is not suitable as a good passband filter within this frequency range.

\section{Conclusions}
This model is very useful to find the geometric parameters necessary to design a metasurface made of csrr resonators that completely transmits at a fixed frequency within the microwave range. Besides, it was demonstrated by full wave simulation and waveguide experiments the accuracy in the prediction of the resonance frequency of thick resonators. For instance, for the thick csrr we found that $f_{0}^{sim}=$ 2.82 GHz, $f_{0}^{exp}=$2.82 GHz which tends to $f_{0}^{model}=$2.86 GHz. Besides, it can be verified that the model remains valid for higher order resonances, as shown with the csr3 and its resonance at $f_{1}$ within the same frequency range. In the future, it would be necessary to vary the size of each resonator in order to find the resonance peaks for all the geometries proposed.
  
\end{document}